# Arbitrary helicity control of circularly polarized light from lateral-type spin-polarized light-emitting diodes at room temperature

Nozomi Nishizawa[1, a)], Masaki Aoyama[1], Ronel C. Roca[1], Kazuhiko Nishibayashi[1], and Hiro Munekata[1, b)]

[1]*Institute of Innovative Research, Tokyo Institute of Technology, 4259-J3-15 Nagatsuta, Midori-ku, Yokohama 226-8503, Japan*

**ABSTRACT**

We demonstrate arbitrary helicity control of circularly polarized light (CPL) emitted at room temperature from the cleaved side-facet of a lateral-type spin-polarized light-emitting diode (spin-LED) with two ferromagnetic electrodes in an anti-parallel magnetization configuration. Driving alternate currents through the two electrodes results in polarization switching of CPL with frequencies up to 100 kHz. Furthermore, tuning the current density ratio in the two electrodes enables manipulation of the degree of circular polarization. These results demonstrate arbitrary electrical control of polarization with high speed, which is required for the practical use of lateral-type spin-LEDs as monolithic CPL light sources.

(97/100 words)

---

a) Electronic mail: nishizawa.n.ab@m.titech.ac.jp and b) mail: munekata.h.aa@m.titech.ac.jp.

Circularly polarized light (CPL) has unique applicability in various fields beyond the application area of linearly polarized light (LPL). Circularity-related difference in photo-response to right- and left-handed CPL have been used in chemical analyses and for separation of optical isomers [1, 2], while differences in reflectivity of optical orthogonal components have been utilized in ellipsometry [3]. Three-dimensional displays based on right- and left-handed CPL [4] accommodate head tilt and prevent simulator sickness due to crosstalk between helicities, which owes to the rotational symmetry of CPL [5]. In addition to these existing technologies, numerous potential applications based on CPL are proposed and being developed. Entanglement of photon polarization including LPL and CPL can be utilized in quantum cryptographic communication techniques [6, 7]. Because CPL can maintain its polarization longer than LPL when it undergoes many-body scattering on large (larger than the wavelength) particles such as cell nuclei, it can be used as a probe for bio-tissues, [8, 9]. The realization and development of these applications requires a light source with the following characteristics: high polarization emission at room temperature (RT), monolithic device construction without the use of another excitation light source and external electric/magnetic fields, and high-speed and arbitrary electrical controllability of polarization, as well as compactness for achieving a high integration.

In addition to optical filters consisting of a linear polarizer (LP) and a quarter-wave plate (QWP) with a conventional light source, various CPL emitters have been studied, including organic light-emitting diodes (OLED) with chiral polymers [10], chiral photonic crystal (metamaterials) devices [11, 12], chiral light-emitting transistors [13], spin-polarized vertical-cavity surface-emitting lasers (spin-VCSELs) [14 - 17], and spin-polarized light-emitting diodes (spin-LEDs) [18, 19]. All these devices have their advantages and disadvantages for future applications. The use of optical polarizing filters requires mechanical rotation to change the polarization. OLED with chiral polymers and chiral photonic crystals can emit CPL with high degrees of polarization and intensities, while the polarity is in general fixed as an intrinsic property of the constituent materials and structures. Using metamaterials with Archimedean spiral shapes, enantiomeric switching has been demonstrated by selecting the deformation direction by pneumatic force. The switching speed of deformation is defined by the natural frequency of the spiral shapes and is on the order of several kHz [12]. Additionally, electrical controllability of the polarization has been demonstrated in electric-double-layer transistors with monolayers of transition metal dichalcogenides. However, the polarization switching is accompanied by a shift of the emission wavelength, which is one of the outstanding issues for the practical use. Spin-VCSELs have achieved a high degree of polarization and intensity and high-speed modulation, yet the requirement of another excitation light source narrows their application range.

Spin-LEDs had been once considered unsuitable for practical use as light sources because their circular polarization $P_{circ}$ was comparably low at room temperature (RT) due to spin-scattering in a semiconductor and the requirement of applying high external magnetic fields. Indeed, the highest $P_{circ}$ values were reported to be 0.3–0.4 for the surface emission from a GaAs-based spin-LED with an applied magnetic field of $B = 0.8$ T [20]. However, we recently reported almost pure CPL emission at RT from the cleaved side-edge of well-designed lateral-type spin-LEDs [21]. Devices of this type do not require an external magnetic field and instead use a remnant magnetization state whose direction is parallel to the emission direction. Furthermore, we demonstrated the polarization controllability at 10 K on the basis of the fact that the helicity of CPL depends on the spin-direction of injected electrons [22].

In this paper, we report the arbitrary helicity control of CPL in spin-LEDs at RT. We fabricate GaAs-based lateral-type spin-LEDs with dual electrode in an antiparallel magnetization configuration. Electrical switching between these two electrodes generates periodically changing polarization, and adjustment of the current flowing into the electrodes enables us to manipulate the polarization arbitrarily.

The structure of the tested spin-LED devices consists of a GaAs-based double heterostructure (DH), a 1-nm-thick γ-phase $AlO_x$ (γ-$AlO_x$) layer [23], and a pair of Fe electrodes, as schematically shown in Figure 1(a). The DH structure was grown on a *p*-GaAs (001) substrate using metal-organic vapor phase epitaxy at Optowell Co., Ltd., which provides high optical quality with fewer luminescence quenching centers causing non-radiative recombination. Regarding the design of the LED structure, there are three notable points. First, the *p*-GaAs active layer is sufficiently thick to have degenerate states in the valence band. In contrast to devices with quantum wells, where the spin direction of holes is restricted along the confinement direction, the structure with a degenerate active region allows CPL emissions in any direction according to orientation of the injected spins. Therefore, a thick active region enables extracting CPL from spins in in-plane remnant magnetization states. Second, the doping in the active region is *p*-type. The radiative recombination time $\tau_{rec}$ in a doped semiconductor ($p$ cm$^{-3}$) can be described as $\tau_{rec} = (Bp)^{-1}$. Here, $B$ is the bimolecular recombination coefficient, $B \sim 1.03 \times 10^{-10}$ cm$^3$ s$^{-1}$ in *p*-GaAs [24]. To achieve steeper switching of polarization, shorter $\tau_{rec}$ is required. In the active layer doped with carbon at $p = 1 \times 10^{18}$ cm$^{-3}$, $\tau_{rec}$ is estimated to be ~ 10 ns, which enables switching of polarization with rates up to 100 MHz. Finally, the thickness of the *n*-AlGaAs transport (cladding) layer should be chosen appropriately. For light-emitting devices, the active region should be far away from the metal electrodes to reduce the absorption of the emitted light. On the other hand, a thin transport layer is desirable taking into account spin relaxation during the travel of injected electrons. Considering this trade-off, we chose a thickness of 500 nm for the $n$-AlGaAs layer. Therefore, the value for spin polarization of electrons $P_e$ reached at the active region is estimated to be $P_e = 0.24$ through the spin polarization in Fe, $P_{Fe} \sim 0.4$, and the relation $P_e = P_{Fe}\exp(\sqrt{D\tau_s}/L)$, where $D = 98.4$ cm$^2$/s is the diffusion constant [25], $\tau_s = 0.11$ ns is the spin relaxation time [26], and $L = 500$ nm is the thickness of $n$-AlGaAs layer. Less than 5% of the emission energy is absorbed in the metal electrodes [21]. Subsequently, we prepared a $\gamma$-AlO$_x$ layer with a thickness of 1 nm for sufficient spin injection into the LED structure. An ultra-thin crystalline AlO$_x$ layer on GaAs surface can be obtained by static post-oxidation of an epitaxial aluminum layer with a well-controlled thickness. The preparation method is described in detail in ref. [23]. Through the crystallographic analysis of the cross-sectional micrograph with a transmission electron microscope (Figure 1 (b)), the obtained layer is confirmed to have a crystalline structure close to $\gamma$-Al$_2$O$_3$ which corresponds to the low-temperature phase of Al$_2$O$_3$ crystals. The spin injection junctions with a structure of Fe/1 nm-thick $\gamma$-AlO$_x$ layer/GaAs shows a relatively high spin injection efficiency $\eta$ at 10K, $\eta \sim 0.63$ [23]. In this junction, the density of interface states $D_{it}$ at the $\gamma$-AlO$_x$ layer/GaAs interface estimated through the admittance spectra obtained in capacitance–voltage measurements is $D_{it} \sim 7 \times 10^{11}$ cm$^{-2}$ eV$^{-1}$, which is 2–3 orders of magnitude less than the typical values for amorphous-AlO$_x$/GaAs junctions. The $\gamma$-AlO$_x$ layer also acts as a protection layer to prevent the diffusion of metal ions into the semiconductor layer, as an electrical equipotential layer to achieve electrical uniformity at the semiconductor surface, and as a robust insulator layer against high current densities to maintain tunnel injection up to several hundred A/cm$^2$. Finally, a pair of Fe electrodes were deposited by electron beam evaporation on the γ-AlO$_x$ layer and fabricated by photolithography as two 40-μm stripes with a spatial separation of 250 μm. The two electrodes have different thicknesses, 100 and 30 nm, providing a sufficiently large difference in the switching fields to achieve an anti-parallel magnetization configuration due to the shape anisotropy, as shown with the *M–H* curves in Figure 1 (c). The manipulation of the external magnetic field results in an antiparallel magnetization configuration in remnant states; we first apply an external field of $H = +5$ kOe along the <1–10> axis to achieve a parallel configuration, then decrease the field to $H = -30$ Oe to reverse only the magnetization of the 100-nm-thick electrode, and finally remove the external field. The obtained antiparallel configuration is stable even without the external magnetic field because the lines of magnetic force of the pair of magnetic electrodes make a closed loop. The voltage dependences of current densities in the 100-nm and 30-nm electrodes of the fabricated devices, $J_{100}$ and $J_{30}$, respectively, show almost identical rectification curves with a slight difference in resistance, as shown in Figure 1(d). Illustrated in Figure 1(e) is the optical setup to detect the helicity-dependent components of electroluminescence (EL) emitted from the edge of the device. The EL spectra of the σ$^+$/σ$^-$ components were detected with a photomultiplier tube (PMT) through an LP with a fixed angle of 45° and a QWP with an angle of 0°/90°, respectively. The time traces of the σ+/σ− EL intensities were measured with a digital oscilloscope. Through the synchronous processing of each component detected individually, the oscillations of polarization were evaluated.

      Figure 2 shows the helicity-dependent EL spectra for continuous current injection into the (a) 100-nm and (b) 30-nm electrodes with current densities of $J_{100} = 4.0$

A/cm² at $V$ = 2.8 V and $J_{30}$ = 3.8 A/cm² at $V$ = 3.0 V, respectively. The insets are photographs of the device with EL emissions, which show that bright areas of EL emission are identical and uniform over the entire device. These EL spectra show almost identical peaks at 1.428 eV and a negligible difference in spectral shapes between the electrodes and helicities, which indicates that the detected emission signals come from the common radiative recombination and propagation processes in the LED structure. Whereas the intensities of the majority/minority components are almost the same between the electrodes, the polarities are the opposite according to the antiparallel magnetization of the electrodes. The degrees of circular polarization $P_{circ}$ are $P_{circ}$ = +10 ~ +12 % and $P_{circ}^{30}$ = –10 ~ –12 % for the 100-nm and 30-nm electrodes, respectively. Figure 2(c) shows the current density dependence of $P_{circ}$ and total EL intensity $I$, with the superscripts of 100 and 30 denoting the respective electrodes. The $P_{circ}$ values are almost constant within the range of current density $J$ from 2.0 to 3.8 A/cm² while $I^{100}$ and $I^{30}$ show monotonically increasing behavior. The helicity switching and tuning experiments described below were carried out in this current density region. An increase of the current density beyond this region induces a gradual decrease in $P_{circ}$, and a further increase above 100 A/cm² causes a non-linear decrease in the minority components of EL intensities and consequently, saturation of $P_{circ}$ up to ~ 100 %, as shown in ref. [21]. The control experiments where the γ-AlO$_x$ layer is replaced by a Schottky barrier layer, n⁺-GaAs with a thickness of 15nm and $n$ ~ 8 × 10¹⁸ cm⁻³, show no evidence of spin-dependent radiative recombination, although the spectral shapes are almost the same.

      Electrical polarization switching was implemented by sending square current waves into the two electrodes with a two-channel current source (Figure 1(e)). The phases of the square current waves were different from each other by a half period. The frequency

of the square current waves, $f$, was varied from 1-kHz to 100 kHz. The current densities were chosen as 3.3 and 3.2 A/cm$^2$ for the 100-nm and 30-nm electrodes, respectively, in order to obtain similar EL intensities. Figure 3 shows the experimental results for $f =$ (a) 1 k, (b) 10 k, and (c) 100 kHz. The red and blue lines shown in the two upper rows in Figure 3 represent the time-dependent intensities of the σ+ and σ− EL components ($I(σ+)$ and $I(σ−)$, respectively), and the green lines shown in the third row represent the temporal profiles of the total EL intensities calculated as a sum of $I(σ+)$ and $I(σ−)$. Although these data were collected independently, we can plot the data on the same time axis using appropriate synchronous processing in the digital oscilloscope. For all three frequencies indicated in the figure, distinct switching of helicity-dependent components is observed, while the periodic changes in the total EL intensities are small. The σ+ and σ− components oscillate out of phase. The time traces of $P_{circ}$ calculated as $P_{circ} = \{I(σ+) − I(σ−)\}/\{I(σ+) + I(σ−)\}$, are represented by the black line in the bottom row in Figure 3. Periodic inversion of $P_{circ}$ values is observed across zero polarization according to the frequency of electrical signals, $f$. With increasing frequency, the squareness of the oscillations is somewhat reduced. We suppose that, when $1/f$ approaches $\tau_{rec}$, the spins injected from different electrodes coexist in the active layer at the switching moment. The superposition of emission signals due to the spin injection form different electrodes is also inferred from the time profile of the total intensity with larger noise, with a typical plot shown in Figure 3(c). Only taking account of $\tau_{rec}$, which is on the order of 10 ns at RT, switching frequencies up to sub-GHz would be possible.

Further, we demonstrate arbitrary modulation of polarization at RT. For the current densities flowing into the 100-nm and 30-nm electrodes in the range of $2.3 < J_{100}$ and $J_{30} < 3.3$ A/cm$^2$, the $P_{circ}$ values do not exhibit large variations and $I$ values vary almost linearly (Figure 2(c)). Therefore, within this current density region, the emission intensity ratio, $I_{100}/(I_{100}+I_{30})$, can be adjusted in a straightforward way keeping the total emission intensities constant. The experimental results are shown in Figure 4. The horizontal axis represents the emission intensity ratio $I_{100}/(I_{100}+I_{30})$. The values of 0 and 1 on the horizontal axis correspond to the cases where current flows into only the 100-nm or 30-nm Fe electrode, respectively. The left vertical axis shows the resultant $P_{circ}$ values calculated from helicity-dependent EL spectra and the right axis indicates the total EL intensities. The EL spectra are shown in the inset figures at specific intensity ratios of $I_{100}/(I_{100}+I_{30}) = 0$, 0.5, and 1.0, from the left to the right. The $P_{circ}$ values can be controlled continuously from negative through zero to positive values by tuning the currents sent to each electrode. At $I_{100}/(I_{100}+I_{30}) = 0.5$, both of the EL components are overlapped with the same intensities, so that $P_{circ}$ is almost zero. One may suppose that the polarization state at $P_{circ} = 0$ is a linear polarization along the horizontal direction, which corresponds to the propagating mode in a rectangular waveguide. However, the $P_{circ} = 0$ state is not a specific linear polarization which is verified experimentally by rotating of a linear polarizer. For current densities smaller than 3.3 A/cm$^2$, the stable modulation of polarization is possible. Increasing the current density further gives rise to the deviation of minority components of the EL intensity from a monotonic increase. In this region, the resultant $P_{circ}$ values are unstable against slight electrical changes, which impairs the arbitrary controllability of polarization. However, from another point of view, systematic investigation of this situation where the minority spins are injected into the process cycles with positive feedback for majority spins could provide insight into the physical mechanism of the saturation of circular polarization [21]. Recall, in the helicity switching experiments at higher frequency ($f \geq 500$ kHz), CPL emissions become unstable. This effect may also be associated with coexistence of two opposite spins in the active region at switching period.

In summary, lateral-type spin-LED devices with dual ferromagnetic electrode were prepared. A pair of electrodes were arranged to achieve anti-parallel magnetization. Alternate driving currents were injected into the two electrodes to demonstrate clear polarization switching with frequencies up to 100 kHz. Tuning the current density ratio enables emission with arbitrary control of the degree of circular polarization in the range from –11 to 9 %. We demonstrated two functionalities required for practical use at RT. Along with pure CPL emission [21], this indicates the superiority of spin-LED devices to various other CPL emitters.

(2664 / ~2700 words)

**Acknowledgements**

The authors would like to acknowledge support in part by the Advanced Photon Science Alliance from the Ministry of Education, Culture, Sports, Science and Technology (MEXT). This work was partly supported by the Ministry of Internal Affairs and Communications (MIC) through the Strategic Information and Communications R&D Promotion Programme (SCOPE) No. 162103004, a Grant-in-Aid for Scientific Research (No. 17K14104) from the Japan Society for Promotion of Science (JSPS), the Matching Planner program from Japan Science and Technology Agency (JST), and the Murata Science Foundation.

.

(26 / 20 references)

Figure caption

Figure 1

(a) Schematic cross-sectional device structure of tested spin-LED devices. (b) Cross-sectional TEM image around the $\gamma$-AlO$_x$/$n$-GaAs interface of the device. (c) $J$–$V$ curves of the tested chip with a 100-nm (black) and a 30-nm (red) Fe electrodes. The forward bias is in the direction where electrons are injected from the magnetic electrode into the semiconductor layers. (d) $M$–$H$ curve for a chip with a pair of electrodes. (e) Schematic experimental setup for EL measurements. Two electrodes are connected to a two-channel current source that can apply square current waves and modulate the current densities independently. Light emitted from the cleaved side-facet of the device passes through a quarter-wave plate (QWP) and a linear polarizer (LP) and is collected by a cooled photo multiplier tube. The fast axis of the QWP is set at 0° and 90° to detect the σ+ and σ− components, respectively, while the optical axis of the LP is fixed at 45°. The detected signals are synchronized with the current source signals with a digital oscilloscope connected with the current source through a reference line.

Figure 2

Helicity-specific EL spectra obtained at RT from the cleaved side-facet of a spin-LED with a pair of electrodes by sending currents to (a) 100-nm and (b) 30-nm electrode, respectively. Current densities are (a) $J = 4.0$ A/cm$^2$ at $V = 2.8$ V and (b) $J = 3.8$ A/cm$^2$ at $V = 3.0$ V. The inset figures show far-field images of the device with EL. (c) Current density dependence of circular polarization $P_{\text{circ}}$ (red) and total EL intensity $I$ (blue). The closed and open symbols and lines represent the data obtained by sending currents to the 100-nm and 30-nm electrodes, respectively.

Figure 3

Experimental data for electrical polarization switching by sending square current waves into two electrodes with frequencies $f =$ (a) 1 k, (b) 10 k, and (c) 100 kHz at RT. Red and blue lines in the top two rows represent the $\sigma+$ and $\sigma-$ components, respectively. Green lines in the third row represent the total EL intensities. Black lines show the time traces of the $P_{circ}$ values calculated as $P_{circ} = \{I(\sigma+) - I(\sigma-)\}/\{I(\sigma+) + I(\sigma-)\}$.

Figure 4

Experimental data for arbitrary modulation of polarization at RT. The horizontal axis shows the EL intensity ratio $I_{100}/(I_{100}+I_{30})$. Red symbols and line show the $P_{circ}$ values on the left vertical axis, and blue color corresponds to the total EL intensity indicated on the right axis. The plots in the upper inset show helicity-dependent EL spectra at specific intensity ratios of $I_{100}/(I_{100}+I_{30}) = 0$, 0.5, and 1.0, from the left to the right, respectively.

Figure

Figure 1

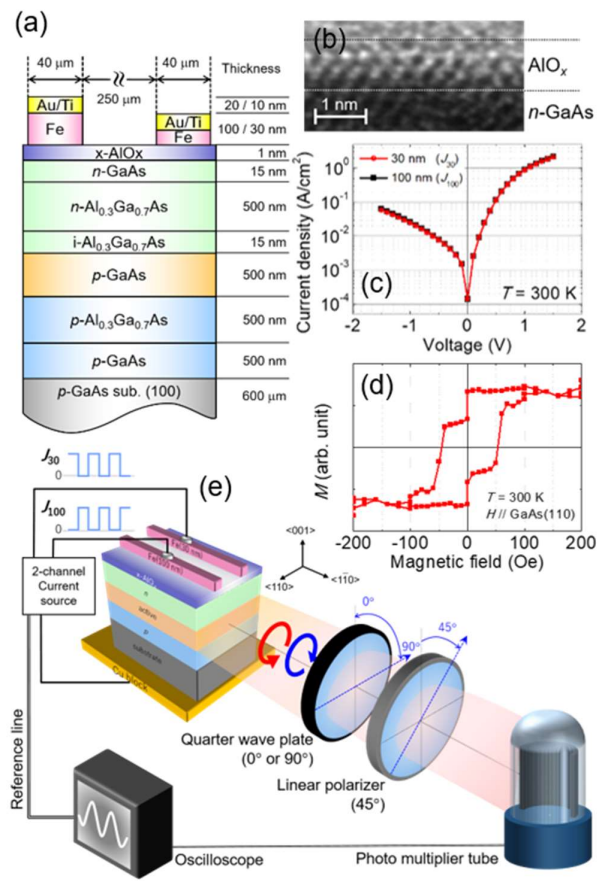

N. Nishizawa *et al.*

Figure 2

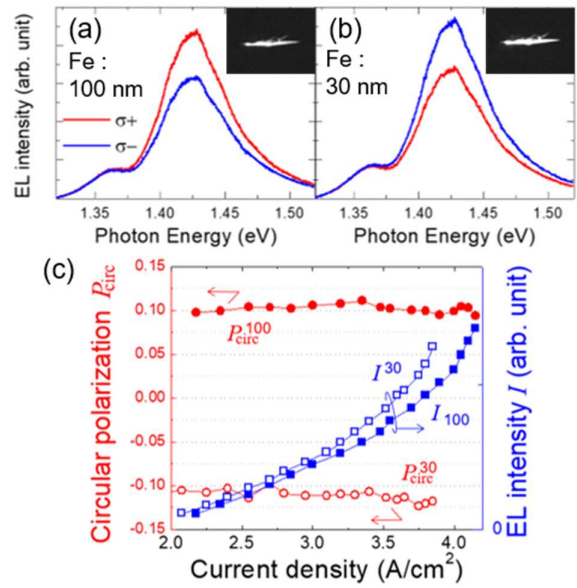

N. Nishizawa *et al.*,

Figure 3 (2-column figure)

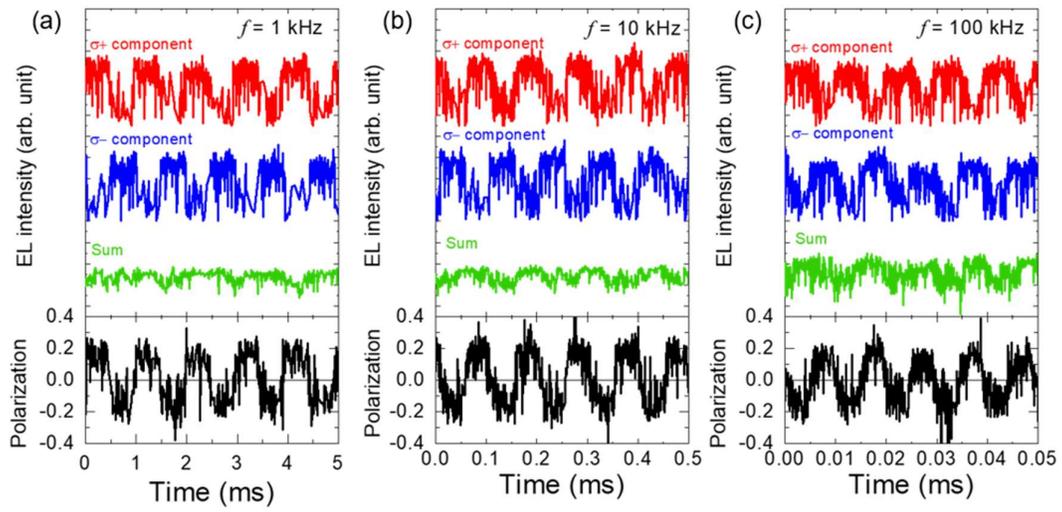

N. Nishizawa *et al.,*

Figure 4

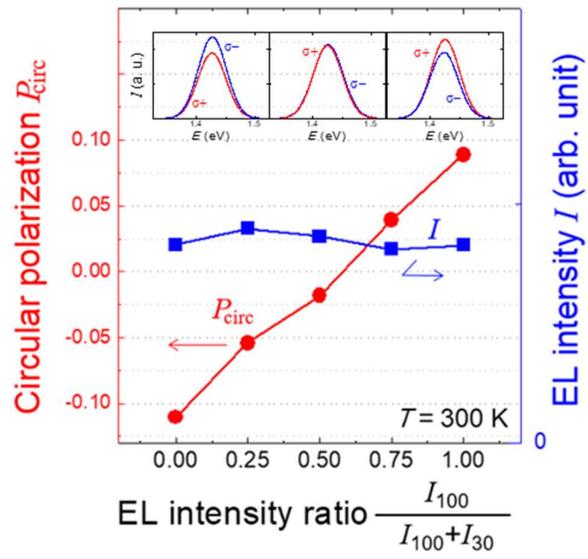

N. Nishizawa *et al.*,